\documentclass[twocolumn,showpacs,prl]{revtex4}

\usepackage{graphicx}
\usepackage{color}\usepackage{enumerate}
\usepackage{amsmath}\usepackage{amssymb}\usepackage{stmaryrd}
\usepackage{multirow}
\usepackage{float}
\usepackage{longtable,booktabs}
\usepackage{hyperref}\usepackage{url}
\begin{document}

\title{Braiding Statistics and Congruent Invariance of Twist Defects in Bosonic Bilayer Fractional Quantum Hall States}

\author{Jeffrey C.Y. Teo}\email{cteo@illinois.edu}
\author{Abhishek Roy}
\author{Xiao Chen}
\affiliation{Department of Physics, Institute for Condensed Matter Theory, University of Illinois at Urbana-Champaign, IL 61801, USA}

\begin{abstract}
We describe the braiding statistics of topological twist defects in abelian bosonic bilayer $(mmn)$ fractional quantum Hall (FQH) states, which reduce to the $\mathbb{Z}_n$ toric code when $m=0$. Twist defects carry non-abelian fractional Majorana-like characteristics. We propose local statistical measurements that distinguish the fractional charge, or species, of a defect-quasiparticle composite. Degenerate ground states and basis transformations of a multi-defect system are characterized by a consistent set of fusion properties. Non-abelian unitary exchange operations are determined using half braids between defects, and projectively represent the sphere braid group in a closed system. Defect spin statistics are modified by equating exchange with $4\pi$ rotation. The braiding $S$ matrix is identified with a Dehn twist (instead of a $\pi/2$ rotation) on a torus decorated with a non-trivial twofold branch cut, and represents the congruent subgroup $\Gamma_0(2)$ of modular transformations.
\end{abstract}

\maketitle

Exotic anyonic excitations~\cite{ArovasSchriefferWilczek84, Wilczekbook}, such as Ising and Fibonacci anyon, in non-abelian topological phases~\cite{MooreRead, GreiterWenWilczek91, NayakWilczek96, Volovik99, ReadRezayi, ReadGreen, Ivanov, Kitaevchain, SlingerlandBais01, Kitaev06} have promising implications in topological quantum computing~\cite{Kitaev97, OgburnPreskill99, Preskilllecturenotes, FreedmanKitaevLarsenWang01, ChetanSimonSternFreedmanDasSarma, Wangbook} due to their non-abelian braiding properties and ability to store quantum information non-locally. More recently topological defects that carry fractional Majorana-like characteristics are theoretically predicted to be present in abelian topological systems such as fractional topological insulator - ferrormagnet - superconductor heterostructures~\cite{FuKane08, LindnerBergRefaelStern, ClarkeAliceaKirill, MChen, Vaezi}, and dislocations in abelian bilayer fractional quantum Hall states~\cite{BarkeshliQi, BarkeshliQi13}. They can be conceptually studied as twist defects in symmetry enchanced exact solvable spin/rotor models like the Kitaev toric code~\cite{Kitaev97, Kitaev06, Bombin, KitaevKong12}, the $\mathbb{Z}_n$ Wen plaquette model~\cite{Wenplaquettemodel, YouWen, YouJianWen}, the Bombin-Martin color code~\cite{BombinMartin06, Bombin11} and its $\mathbb{Z}_n$ generalization~\cite{TeoRoyChen13long}.

Twist defects are semiclassical point-like objects that violate an ungauged anyonic symmetry of the system by twisting the anyon label of a circling quasiparticle by the symmetry operation, such as the electric-magnetic duality of the toric code or bilayer symmetry in the $(mmn)$-FQH state. Unlike quantum deconfined anyons in a topological phase, twist defects are not excitations of a quantum Hamiltonian and there is no superposition between quantum states in different defect configurations. The semiclassical distinction of defects from quantum anyons manifest in braiding behavior, and we demonstrate this by twofold twist defects in abelian $(mmn)$-FQH states~\cite{Wentopologicalorder90, WenchiralLL90, WenZee92, Wenbook, Fradkinbook, BoebingerJiangPfeifferWest90, SuenJoSantosEngelHwangShayegan91, EisensteinBoebingerPfeifferWestHe92} with a $\mathbb{Z}_2$ bilayer or spin symmetry. This includes the $\mathbb{Z}_n$ toric code as a special case for $m=0$.

An abelian quantum Hall state with bilayer symmetry is described by the effective topological field theory $\mathcal{L}=K_{IJ}a_I\wedge da_J/4\pi$ characterized by the symmetric integral coefficient $K$-matrix \begin{align}K=\left(\begin{array}{*{20}c}m&n\\n&m\end{array}\right)\label{Kmatrix}\end{align} and the charge vector ${\bf t}=(1,1)$. The theory is invariant under the twofold bilayer symmetry represented by the Pauli matrix $\sigma_x$, and consists of bosonic fundamental charged particles when the diagonal entry $m$ is even. It has a ground state degeneracy of $|\det(K)|=|m^2-n^2|>1$ on a torus. The quasiparticles (up to fundamental bosons) are labeled by two dimensional integral vector ${\bf a}$ on a periodic lattice $\mathcal{A}=\mathbb{Z}^2/K$, and fusion rules are set by lattice addition $[{\bf a}]\times[{\bf b}]=[{\bf a}+{\bf b}]$ modulo $K$. 

Bilayer defects in this letter are demonstrated in one of the following two classes of quantum Hall states to avoid algebraic complications, however we expect the results and arguments apply to the general case. (i) $m=0$. The theory has the identical anyon structure as the $\mathbb{Z}_n$ Kitaev toric code, where quasiparticles are generated by ${\bf e}=(1,0)$ and ${\bf m}=(0,1)$ with braiding characterized by the exchange phases $R^{{\bf e}{\bf e}}=R^{{\bf m}{\bf m}}=R^{{\bf e}{\bf m}}=1$ and $R^{{\bf m}{\bf e}}=e^{2\pi i/n}$. (ii) $n$ is odd. Quasiparticles ${\bf a}=x{\bf t}+\alpha{\bf s}$ are composition of the charge sector ${\bf t}$ and neutral ``spin" sector ${\bf s}=(1,-1)$ with exchange phases $\theta_{\bf t}=R^{{\bf t}{\bf t}}=e^{\frac{2\pi i}{m+n}}$, $\theta_{\bf s}=R^{{\bf s}{\bf s}}=e^{\frac{2\pi i}{m-n}}$, and $R^{{\bf s}{\bf t}}=R^{{\bf t}{\bf s}}=1$. In either of the two cases, the quasiparticle lattice splits into $\mathcal{A}=\mathbb{Z}_{m+n}\oplus\mathbb{Z}_{m-n}$ so that there are unique decomposition of anyons by ${\bf e},{\bf m}$ or ${\bf t},{\bf s}$. The general exchange phase $R^{{\bf a}{\bf b}}$ can be chosen to be multilinear in its two entries so that the hexagon identities of the ribbon category are satisfied with the trivial $6j$-symbols $F^{{\bf a}{\bf b}{\bf c}}_{{\bf a}+{\bf b}+{\bf c}}=1$, which gives trivial bootstraps and bendings for quasiparticle string-nets. When $n$ is odd, this gauge is facilitated by the {\em invertibility} of 2 modulo $m^2-n^2$ and we denote the abbreviation $2^{-1}$ for the integer $(m^2-n^2+1)/2$ while $1/2$ means the regular fraction $0.5$. For instance $R^{{\bf a}{\bf b}}=e^{2\pi i2^{-1}{\bf a}^TK^{-1}{\bf b}}=\exp\left(2\pi i\frac{m^2-n^2+1}{2}{\bf a}^TK^{-1}{\bf b}\right)$ is the exchange phase in case (ii).

The twisting action of a defect on an encircling quasiparticle is given by the twofold symmetry operation ${\bf a}\to\sigma_x{\bf a}$ that interchanges ${\bf e}\leftrightarrow{\bf m}$ for $\mathbb{Z}_n$ toric code or layer/spin conjugation ${\bf s}\to-{\bf s}$ for $(mmn)$-states. The bilayer symmetry leaves the ${\bf t}$ vector and hence the charge of a quasiparticle invariant. Since charge is fractionalized in units of $\frac{1}{m+n}$, there are $|m+n|$ species of twofold defects, denoted by $\sigma_\lambda$ for $\lambda=0,\ldots,|m+n|-1$, distinguished by the fractional charge they carry. Species mutation is driven by absorbing or emitting quasiparticles. Fusion or splitting outcome should be insensitive to the detail of how many times a quasiparticle circles around the defect during the absorption or emission process. Defect species are therefore topological quantities that would change by the same amount upon fusion with the quasiparticle ${\bf a}$ or its symmetry partner $\sigma_x{\bf a}$. Species label lives in the cyclic quotient group $\mathcal{A}/(1-\sigma_x)\mathcal{A}=\mathbb{Z}_{m+n}$, which is generated by the charge vector ${\bf t}$ and coincides with the quasiparticle charge sector. 

\begin{figure}[ht]
\includegraphics[width=3.3in]{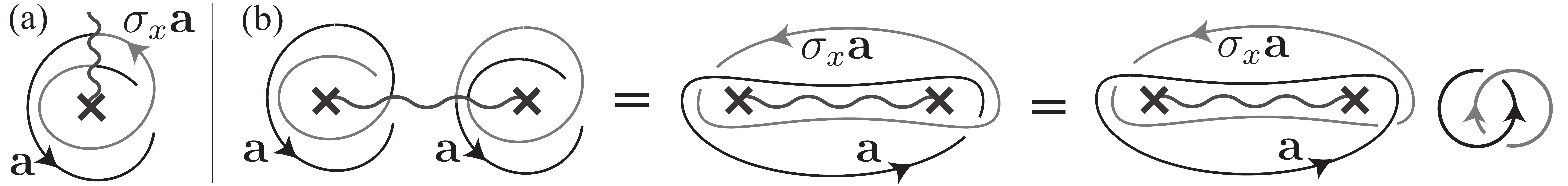}
\caption{(a) The double Wilson loop operator $\hat{\Theta}_{\bf a}$ by moving a quasiparticle ${\bf a}$ twice around a twofold defect (cross). (b) A linking phase $e^{2\pi i{\bf a}^TK^{-1}\sigma_x{\bf a}}$ from joining two double Wilson operators $\hat{\Theta}_a$ around a pair of defects. The two unlinked ${\bf a}$ and $\sigma_x{\bf a}$ loops can then be absorbed in the condensate if the defects fuse to the vacuum.}\label{fig:doubleloop}
\end{figure}
It should be noticed that the distinction of defect species is not a consequence of $U(1)$ charge symmetry but rather the twofold symmetry of the anyonic system~\cite{TeoRoyChen13long}. Defect species can be measured by the braiding phase accumulated by moving a quasiparticle ${\bf a}$ twice around a twofold defect $\sigma_\lambda$. The quasiparticle trajectory leaves behind a Wilson string operator $\hat{\Theta}_{\bf a}$ depicted in figure~\ref{fig:doubleloop}. As the size of $\hat{\Theta}_{\bf a}$ could be arbitrary small, it does not intersect with any non-trivial Wilson operators in the system and corresponds the good quantum number \begin{align}\Theta^\lambda_{\bf a}=e^{2\pi i{\bf a}^TK^{-1}\sigma_x\left[\frac{1}{2}{\bf a}+\left(\lambda+\frac{m+n}{2}\right){\bf t}\right]}\label{doubleloopphase}\end{align} around a defect with species label $\lambda$. The eigenvalue \eqref{doubleloopphase} is uniquely determined by self-intersection $\hat{\Theta}_{{\bf a}+{\bf b}}=e^{2\pi i{\bf a}K^{-1}\sigma_x{\bf b}}\hat{\Theta}_{\bf a}\hat{\Theta}_{\bf b}$, invariance under twofold symmetry $\hat{\Theta}_{\bf a}=\hat{\Theta}_{\sigma_x{\bf a}}$, and a linking phase shown in figure~\ref{fig:doubleloop} that results in the phase $(\Theta^0_{\bf a})^2=e^{2\pi i{\bf a}^TK^{-1}\sigma_x{\bf a}}$ around a pair of self-conjugate twofold defects $\sigma_0$.

The fusion rules~\cite{Kitaev06, Turaevbook, BakalovKirillovlecturenotes, Wangbook} between twofold defects and quasiparticles conserve charge and are given by \begin{align}{\bf a}\times\sigma_\lambda=\sigma_{\lambda+{\bf a}\cdot{\bf t}},\quad\sigma_{\lambda_2}\times\sigma_{\lambda_1}=\sum_{{\bf a}\cdot{\bf t}=\lambda_1+\lambda_2}{\bf a}\label{defectdefectfusion}\end{align} There are $|m-n|$ charge conserving abelian quasiparticle fusion channels in \eqref{defectdefectfusion} and therefore the quantum dimension of defects is given by $d_{\sigma}=\sqrt{|m-n|}$.\cite{BarkeshliJianQi} 

\begin{figure}[ht]
\includegraphics[width=3.3in]{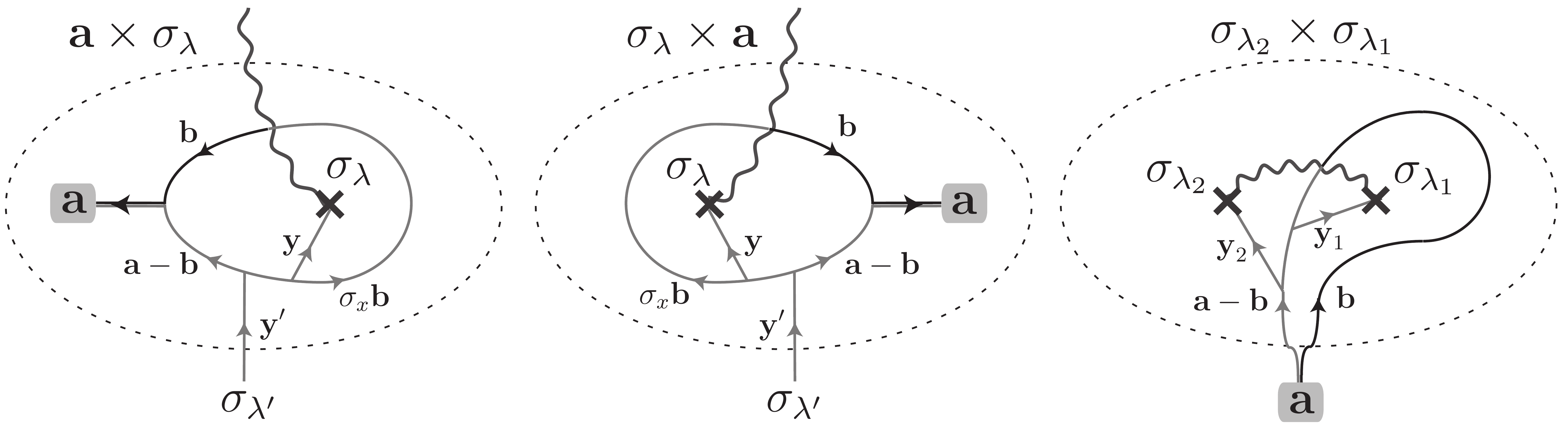}
\caption{Quasiparticle string configurations for splitting states $|V^{\psi_2\psi_1}_{\psi_3}\rangle$. Branch cuts (wavy lines) switch labels ${\bf b}\to\sigma_x{\bf b}$ of passing quasiparticles and end at twofold defects (crosses). (i) For $\mathbb{Z}_n$ toric code where $m=0$, quasiparticle ${\bf a}=a_1{\bf e}+a_2{\bf m}$ is attached with strings ${\bf b}=a_2{\bf m}$ and ${\bf a}-{\bf b}=a_1{\bf e}$, twofold defect $\sigma_\lambda$ is attached with string ${\bf y}=\lambda{\bf e}$. (ii) For $(mmn)$-FQH states with $n$ odd, quasiparticle ${\bf a}$ is attached with strings ${\bf b}={\bf a}-{\bf b}=2^{-1}{\bf a}$, and twofold defect $\sigma_\lambda$ is attached with string ${\bf y}=2^{-1}\lambda{\bf t}$, $2^{-1}=(m^2-n^2+1)/2$.}\label{fig:splittingstates}
\end{figure}
Quantum states of a system of defects (on a closed sphere) are labeled by good quantum numbers represented by admissible fusion channels ${\bf a}_i$ on a fusion tree with defects $\sigma_{\lambda_j}$ as open branches. \begin{align}|{\bf a}_i\rangle=\left|\vcenter{\hbox{\includegraphics[width=1.5in]{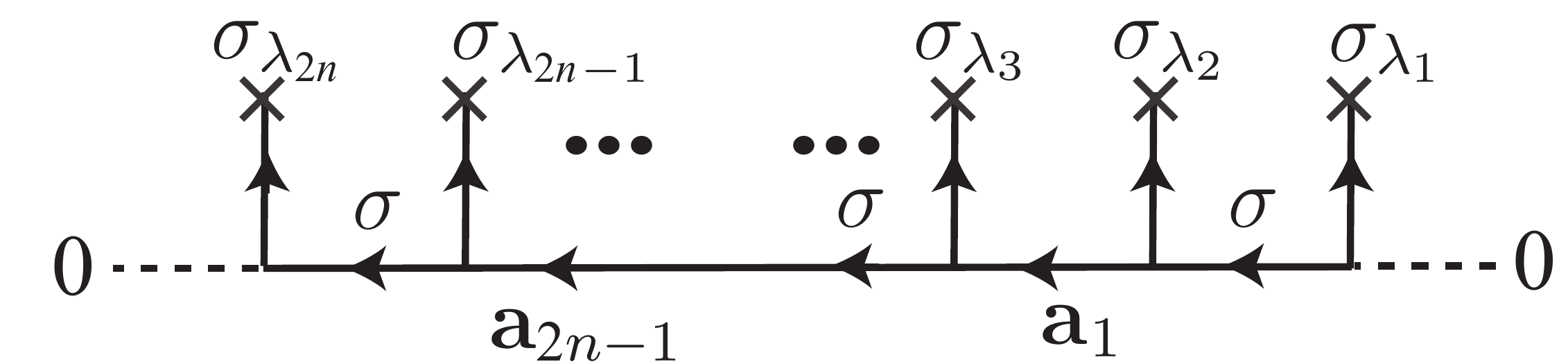}}}\right\rangle\end{align} These form an orthonormal basis of ground states labeled by simultaneous eigenvalues of non-contractible quasiparticle string operators around defects, which form a maximal set of commuting observables. A state is an entangled sum (over boundary conditions) of tensor products of {\em splitting state} $|V^{\psi_2\psi_1}_{\psi_3}\rangle$, one for each trivalent vertex on the fusion tree representing the splitting of object $\psi_3\to\psi_2\times\psi_1$. Splitting states are local states generated by open Wilson strings around the splitting objects, and a gauge is fixed by the particular choice of string configurations shown in figure~\ref{fig:splittingstates}. The overall state can then be constructed by piecing the strings together between local splitting states on the fusion tree by matching boundary conditions.

Basis transformation between different fusion trees are generated by a consistent set of fundamental $F$-moves \begin{align}\left|\vcenter{\hbox{\includegraphics[width=0.5in]{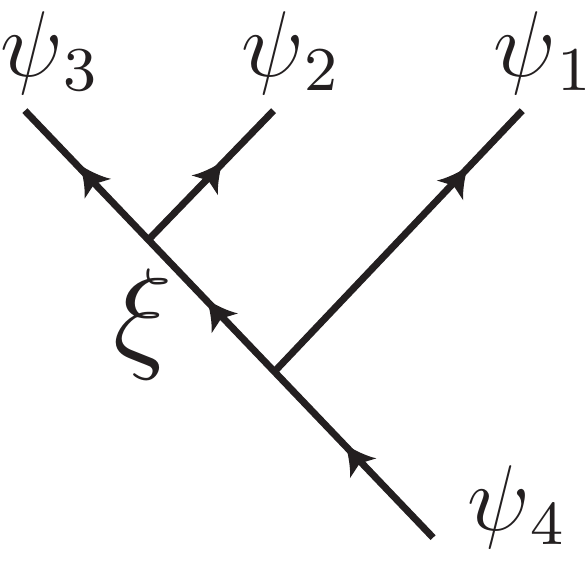}}}\right\rangle=\sum_\eta\left[F^{\psi_3\psi_2\psi_1}_{\psi_4}\right]_\xi^\eta\left|\vcenter{\hbox{\includegraphics[width=0.5in]{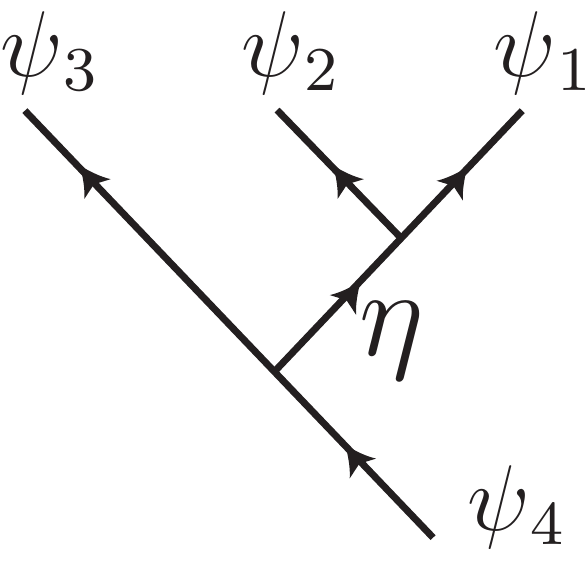}}}\right\rangle\end{align} where the state at the left (and right) hand side is the (entangled sum over boundary conditions of) tensor product of splitting states $|V^{\psi_3\psi_2}_{\xi}\rangle\otimes|V^{\xi\psi_1}_{\psi_4}\rangle$ (resp. $|V^{\psi_2\psi_1}_{\eta}\rangle\otimes|V^{\psi_3\eta}_{\psi_4}\rangle$). These tensor product states can be deformed into each other by crossing, bending and rearranging bootstraps of quasiparticle strings using the $F$ and $R$-symbols of the underlying abelian topological phase. The reults of defect $F$-symbols for $\mathbb{Z}_n$ toric code and $(mmn)$-FQH state are summarized in table~\ref{tab:Fsymbols}.
\begin{table}[ht]
\begin{tabular}{ll}
\multicolumn{2}{c}{$\mathbb{Z}_n$ toric code}\\\hline
$F^{{\bf a}{\bf b}{\bf c}}_{\bf d}$, $F^{{\bf a}{\bf b}\sigma}_{\sigma}$, $F^{\sigma{\bf a}{\bf b}}_{\sigma}$, $F^{{\bf a}\sigma\sigma}_{\bf b}$, $F^{\sigma\sigma{\bf a}}_{\bf b}$ & $1$ \\
$F^{{\bf a}\sigma{\bf b}}_{\sigma}$, $F^{\sigma{\bf a}\sigma}_{\bf b}$ & $e^{-\frac{2\pi i}{n}a_2b_2}$ \\
$\left[F^{\sigma\sigma\sigma}_{\sigma}\right]_{\bf a}^{\bf b}$ & $\frac{1}{\sqrt{n}}e^{\frac{2\pi i}{n}a_2b_2}$
\end{tabular}
\begin{tabular}{ll}
\multicolumn{2}{c}{$(mmn)$-FQH state for $n$ odd}\\\hline
$F^{{\bf a}{\bf b}{\bf c}}_{\bf d}$ & $1$\\
$F^{\sigma\sigma{\bf a}}_{\bf b}$, $F^{{\bf b}\sigma\sigma}_{{\bf b}-{\bf a}}$, & \multirow{2}{*}{$\theta_{\bf s}^{-2^{-1}\alpha\beta}\theta_{\bf t}^{-2^{-1}xy}$}\\
$F^{\sigma{\bf a}{\bf b}}_{\sigma}$, $F^{{\bf a}{\bf b}\sigma}_\sigma$ & \\
$F^{{\bf a}\sigma{\bf b}}_{\sigma}$, $F^{\sigma{\bf b}\sigma}_{{\bf b}+2^{-1}\sigma_x{\bf a}}$ & $\theta_{\bf s}^{\alpha\beta}\theta_{\bf t}^{-xy}$\\
\multirow{2}{*}{$\left[F^{\sigma_{\lambda_3}\sigma_{\lambda_2}\sigma_{\lambda_1}}_{\sigma_{\lambda_1+\lambda_2+\lambda_3}}\right]_{\bf a}^{\bf b}$} & $\frac{1}{\sqrt{|m-n|}}\theta_{\bf s}^{-(\alpha+2^{-1}\beta)\beta}$\\ & $\quad\quad\quad\times\theta_{\bf t}^{(2^{-1})^3(\lambda_1+\lambda_2)(-\lambda_1+\lambda_2+2\lambda_3)}$
\end{tabular}
\caption{Admissible $F$-symbols for twofold defects in (i) $\mathbb{Z}_n$ toric code with quasiparticle decomposition ${\bf a}=a_1{\bf e}+a_2{\bf m}$, ${\bf b}=b_1{\bf e}+b_2{\bf m}$ and (ii) $(mmn)$-FQH states for $n$ odd with quasiparticle decomposition ${\bf a}=x{\bf t}+\alpha{\bf s}$, ${\bf b}=y{\bf t}+\beta{\bf s}$ and $2^{-1}=(m^2-n^2+1)/2$. Here $\theta_{\bf s}=e^{2\pi i/(m-n)}$, $\theta_{\bf t}=e^{2\pi i/(m+n)}$ are the spins for the funamental particles in the charge and spin sectors.}\label{tab:Fsymbols}
\end{table}

Non-abelian defect braiding operations are generated by fundamental exchange permutations of adjacent defects known as $B$-moves. \begin{align}\left|\vcenter{\hbox{\includegraphics[width=0.5in]{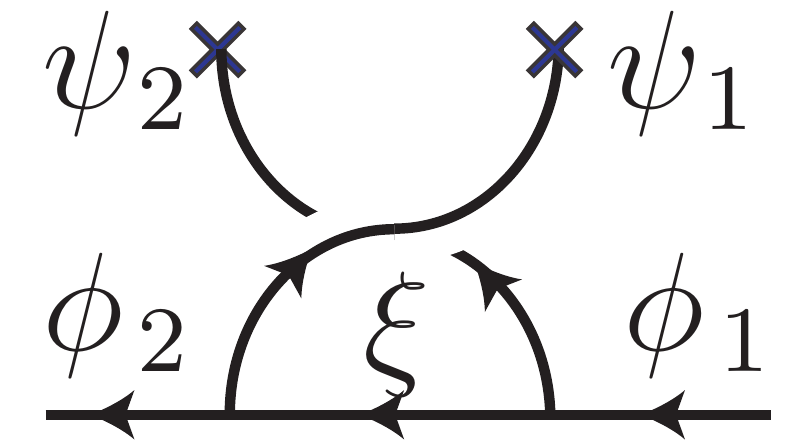}}}\right\rangle=\sum_{\eta}\left[B^{\psi_2\psi_1}_{\phi_2\phi_1}\right]_{\xi}^{\eta}\left|\vcenter{\hbox{\includegraphics[width=0.5in]{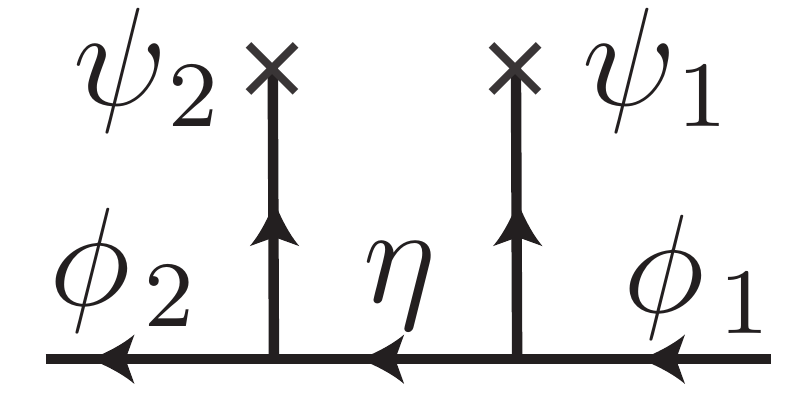}}}\right\rangle\label{Bmatrixdef1}\end{align} 
They can be evaluated by a series of fundamental $F$ and $R$-moves \begin{align}\left[B^{\sigma_2\sigma_1}_{\phi_2\phi_1}\right]_{\xi}^{\eta}&=\sum_{\bf a}\left\{\left[F^{\phi_2\sigma_2\sigma_1}_{\phi_1}\right]_{\eta}^{\bf a}\right\}^\ast R^{\sigma_1\sigma_2}_{\bf a}\left[F^{\phi_2\sigma_1\sigma_2}_{\phi_1}\right]_{\bf a}^{\chi}\end{align} where the $R$-symbol relates the state before and after exchanging a pair of defects $\sigma_2\times\sigma_1$ with fixed fusion outcome ${\bf a}$. It can be derived by twisting the quasiparticle strings of the splitting state $|V^{\sigma_2\sigma_1}_{\bf a}\rangle$ in figure~\ref{fig:splittingstates} by $180^\circ$ and deform back into the untwisted configuration through crossing strings and introducing a double Wilson operator $\hat{\Theta}_{(1/2)\bf a}$. \begin{align}R^{\sigma_{\lambda_1}\sigma_{\lambda_1}}_{\bf a}&=\left\langle\vcenter{\hbox{\includegraphics[width=0.5in]{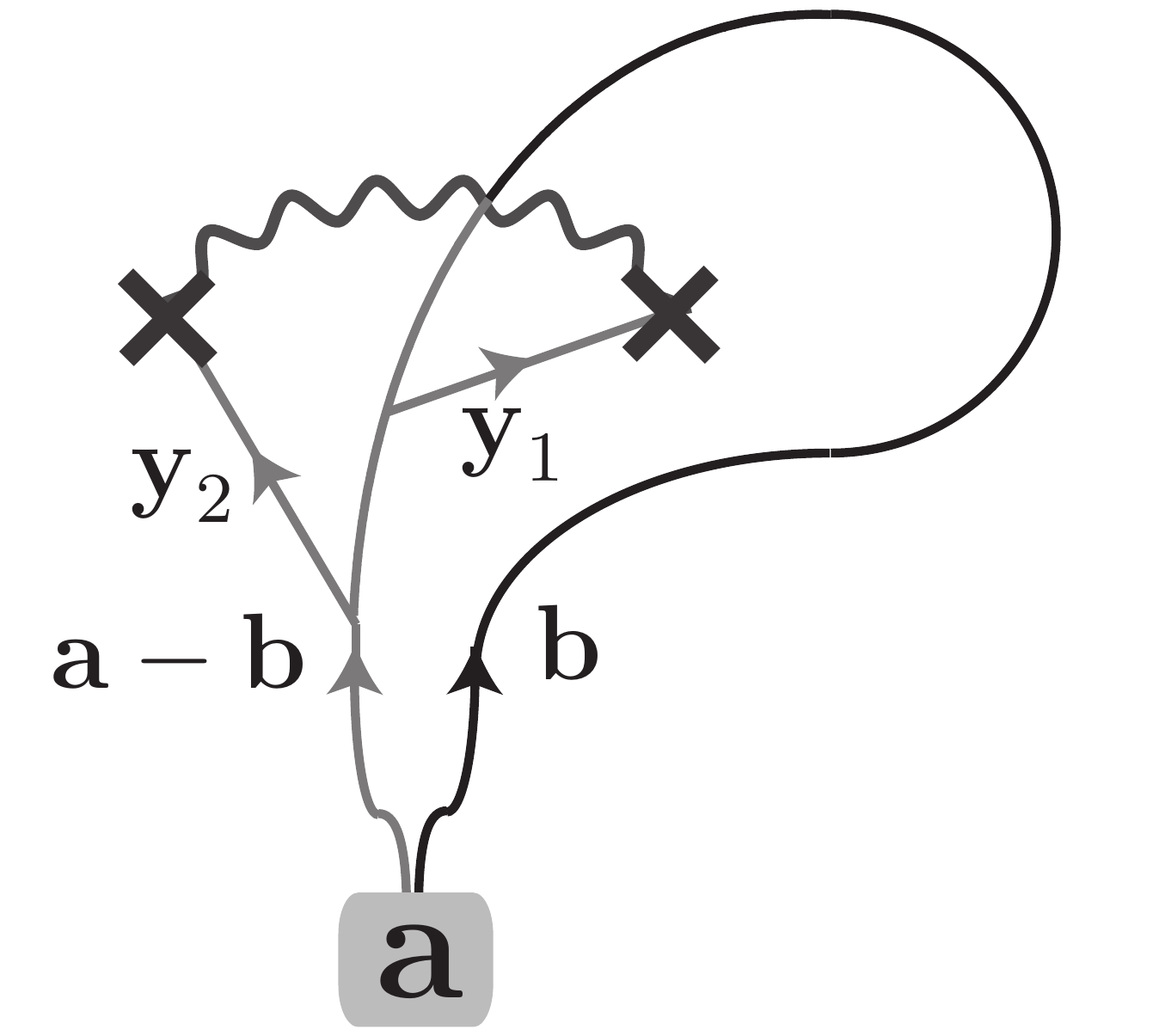}}}\left|\vcenter{\hbox{\includegraphics[width=0.7in]{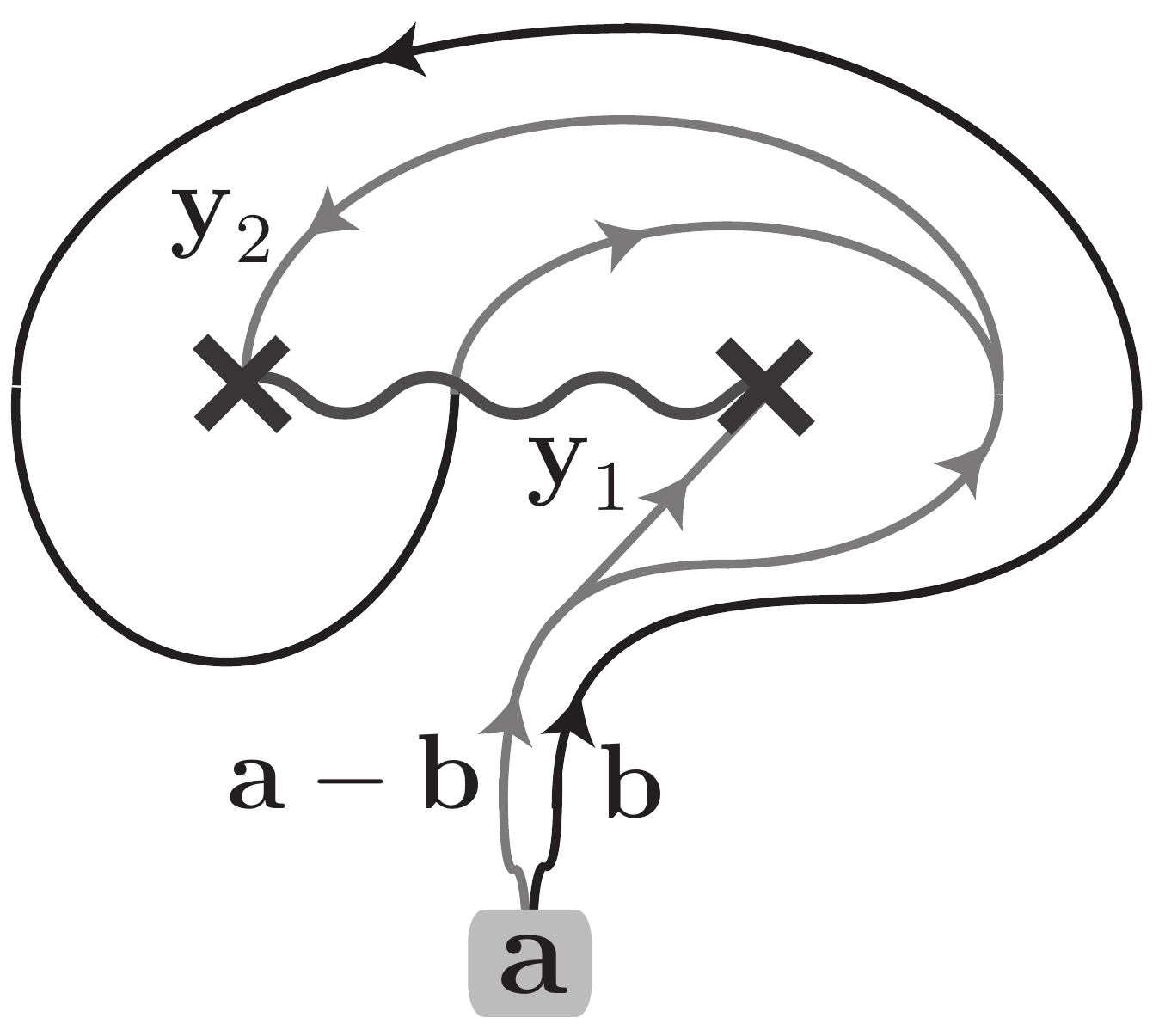}}}\right.\right\rangle\nonumber\\&=\left\{\begin{array}{*{20}c}e^{\frac{2\pi i}{n}a_2\left(\lambda_2+\frac{n}{2}-\frac{1}{2}a_2\right)}\hfill\\\theta_{\bf s}^{2^{-1}\alpha^2}\theta_{\bf t}^{(2^{-1})^3(\lambda_1+\lambda_2)(3\lambda_2-\lambda_1)}\label{Rsymbol}\end{array}\right.\end{align} where the first line applies to $\mathbb{Z}_n$ toric code and the second to $(mmn)$-FQH states with $n$ odd.
\begin{table}[ht]
\begin{tabular}{ll}
\multicolumn{2}{c}{$\mathbb{Z}_n$ toric code}\\\hline\noalign{\smallskip}
$B^{\sigma_{\lambda_2}\sigma_{\lambda_1}}_{{\bf a}{\bf b}}$ & $e^{\frac{2\pi i}{n}(b_2-a_2)\left[\lambda_2+\frac{n}{2}-\frac{1}{2}(b_2-a_2)\right]}$\\
$\left[B^{\sigma_{\lambda_2}\sigma_{\lambda_1}}_{\sigma\sigma}\right]_{\bf a}^{\bf b}$ & $\frac{e^{\frac{i\pi}{4}(n\pm1)}}{\sqrt{|n|}}e^{\frac{2\pi i}{n}(\lambda_2+b_2-a_2)\left[\frac{1}{2}(\lambda_2+b_2-a_2)+\frac{n}{2}\right]}$\\
$\mathcal{M}_{\bf a}^{\bf b}$ & $i^{n\pm1}e^{\frac{2\pi i}{n}\left[\frac{\lambda_1(\lambda_1+n)}{2}+\frac{\lambda_3(\lambda_3+n)}{2}+b_2(\lambda_1+\lambda_4)\right]}\delta_{\bf a}^{\sigma_x{\bf b}}$
\end{tabular}
\begin{tabular}{ll}
\multicolumn{2}{c}{$(mmn)$-FQH states for $n$ odd}\\\hline\noalign{\smallskip}
$B^{\sigma_{\lambda_2}\sigma_{\lambda_1}}_{{\bf a}{\bf b}}$ & $\theta_{\bf s}^{2^{-1}(\alpha-\beta)^2}\theta_{\bf t}^{(2^{-1})^3(\lambda_1+\lambda_2)(3\lambda_2-\lambda_1)}$\\
$\left[B^{\sigma_{\lambda_2}\sigma_{\lambda_1}}_{\sigma\sigma}\right]_{\bf a}^{\bf b}$ & $\frac{e^{\frac{i\pi}{4}(n-m\pm1)}}{\sqrt{|m-n|}}\theta_{\bf s}^{-2^{-1}(\alpha-\beta)^2}\theta_{\bf t}^{(2^{-1})^3(\lambda_1+\lambda_2)^2}$\\
$\mathcal{M}_{\bf a}^{\bf b}$ & $i^{n-m\pm1}\theta_{\bf t}^{(2^{-1})^2(\lambda_1^2+\lambda_2^2+\lambda_3^2+\lambda_4^2)}\delta_{\bf a}^{\sigma_x{\bf b}}$
\end{tabular}
\caption{Fundamental braiding moves, $B$-symbols, of exchanging adjacent defects, and large braid operation $\mathcal{M}$ in \eqref{braidoncearound} for a closed system of four defects. The sign in the $\mathbb{Z}_8$ phases are given by $sgn(m-n)$. }\label{tab:Bsymbols}
\end{table}

The fundamental $B$-moves in table~\ref{tab:Bsymbols} are consistent exchange operations that satisfy the Yang-Baxter equation for adjacent permutations $b^{i}b^{i+1}b^{i}=b^{i+1}b^{i}b^{i+1}$, where $b^{i}$ is the exchange braid between the $(i+1)^{th}$ and $i^{th}$ defect. An exchange is gauge independent whenever it is performed between defects of identical species, and a full braid is always gauge independent. They generate non-abelian unitary operations that {\em projectively} represent the braid group $\mathcal{B}_{2n}(\mathbb{S}^2)$ of $2n$ points on a sphere. Classical braid groups are restricted by a compactification relation if the system lives on a closed sphere, or equivalently the overall fusion channel of the defect system is the vacuum. For instance one expect
the largest braid that moves an object once around all others to be trivial as it should be contractible on the other side of the sphere. \begin{align}\mathcal{M}\equiv b^{1}\ldots b^{N-1}b^{N-1}\ldots b^{1}=\vcenter{\hbox{\includegraphics[width=0.5in]{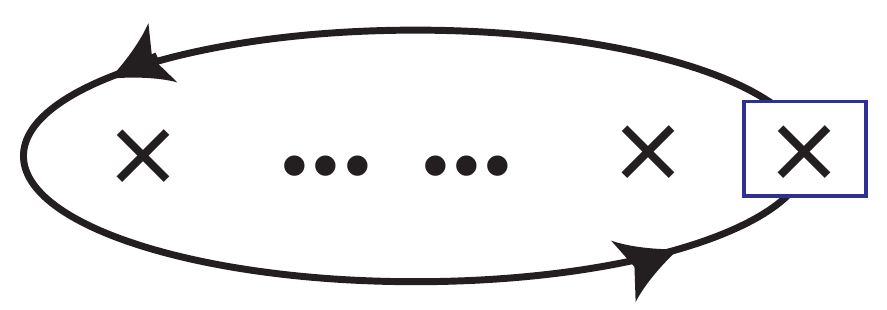}}}=1\label{braidoncearound}\end{align} The violation of this is shown by the non-trivial $\mathcal{M}$ matrices in table~\ref{tab:Bsymbols} in a closed system of four defects with canceling species labels $\lambda_1+\lambda_2+\lambda_3+\lambda_4=0$, where the $|m-n|$-fold ground state degeneracy is labeled by the restricted fusion channel $\overline{\sigma_{\lambda_2}}\times\overline{\sigma_{\lambda_1}},\sigma_{\lambda_4}\times\sigma_{\lambda_3}\to{\bf a}$. 
The large braid operation $\mathcal{M}$ can be interpreted as a flip operator that interchanges the bilayer as the braid, and consequently the twofold branch cut left behind the defect trajectory, is contracted on the other side of the closed sphere. The $\mathcal{M}$-matrix commutes with all braiding operations and generates a center that extends the sphere braid group $\mathcal{B}_{2n}(\mathbb{S}^2)$. It squares to a scalar phase $\mathcal{M}^2=\theta_{\sigma_{\lambda_1}}\theta_{\sigma_{\lambda_2}}\theta_{\sigma_{\lambda_3}}\theta_{\sigma_{\lambda_4}}$ of products of defect exchange spins $\theta_{\sigma_{\lambda}}$ discussed below. $\mathcal{M}$ can thus be viewed as a braiding signature of interchanging bilayer in a defect system.

The exchange matrix \eqref{Rsymbol} defines the spin of a twofold defect by averaging over admissible quasiparticle fusion channel \begin{align}\theta_{\sigma_\lambda}=\frac{1}{d_\sigma}\sum_{\bf a}R^{\sigma_\lambda\sigma_\lambda}_{\bf a}=e^{\frac{i\pi}{4}(n-m\pm1)}e^{\frac{2\pi i}{m+n}\lambda\left(\frac{\lambda+m+n}{2}\right)}\label{exchangespin}\end{align} where the sign in the $\mathbb{Z}_8$ phase is given by $sgn(m-n)$. A defect can be represented as an open end of a twofold branch cut which switches the bilayer, and therefore a $360^\circ$ rotation leaves behind an uncancelable branch cut surrounding the defect. The exchange spin \eqref{exchangespin} however identifies with $720^\circ$ rotation up to the constant $\mathbb{Z}_8$ phase. The defect $\sigma_\lambda$ is equipped with a quasiparticle string ${\bf a}$ that gives the fractional charge $\lambda={\bf a}\cdot{\bf t}$ to the defect. Upon rotation, the string wraps around the defect twice, and can be untwisted by a crossing $\theta_{\bf a}=R^{{\bf a}{\bf a}}$ and absorbing a double Wilson operator $\hat{\Theta}_{\bf a}$ into the condensate. \begin{align}\vcenter{\hbox{\includegraphics[width=0.35in]{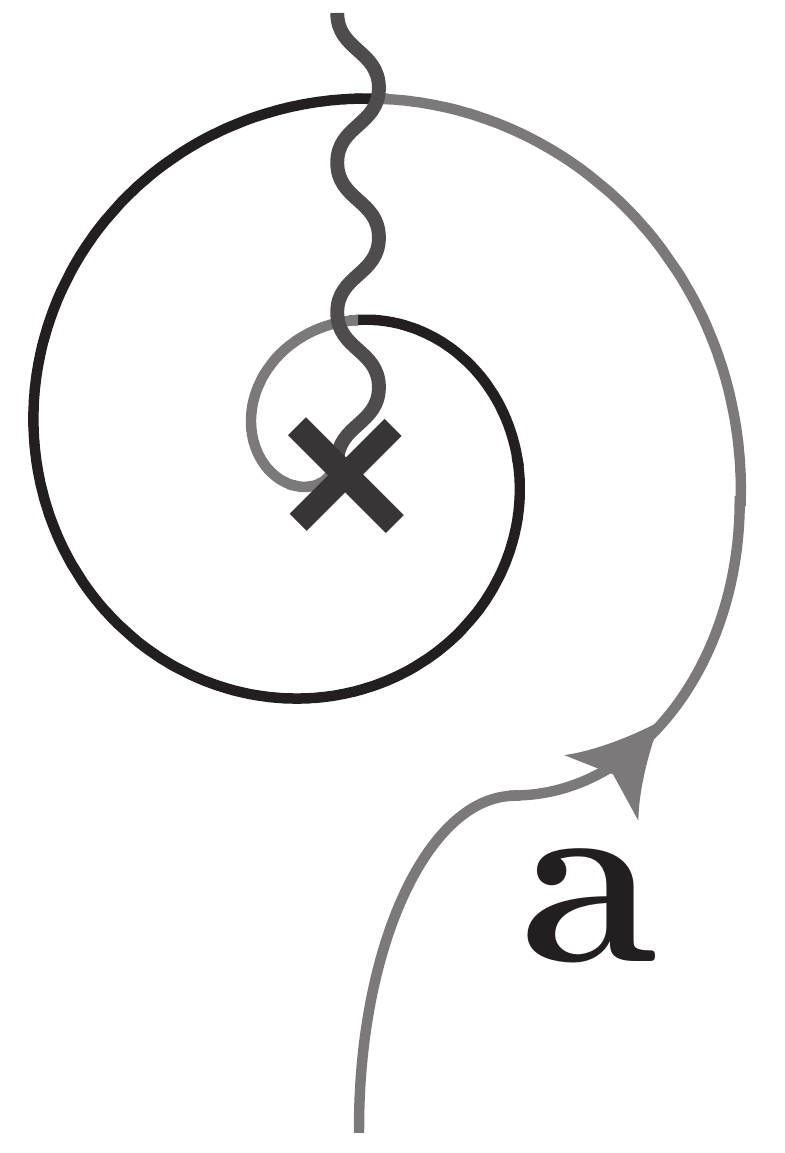}}}&=\theta_{\bf a}\vcenter{\hbox{\includegraphics[width=0.32in]{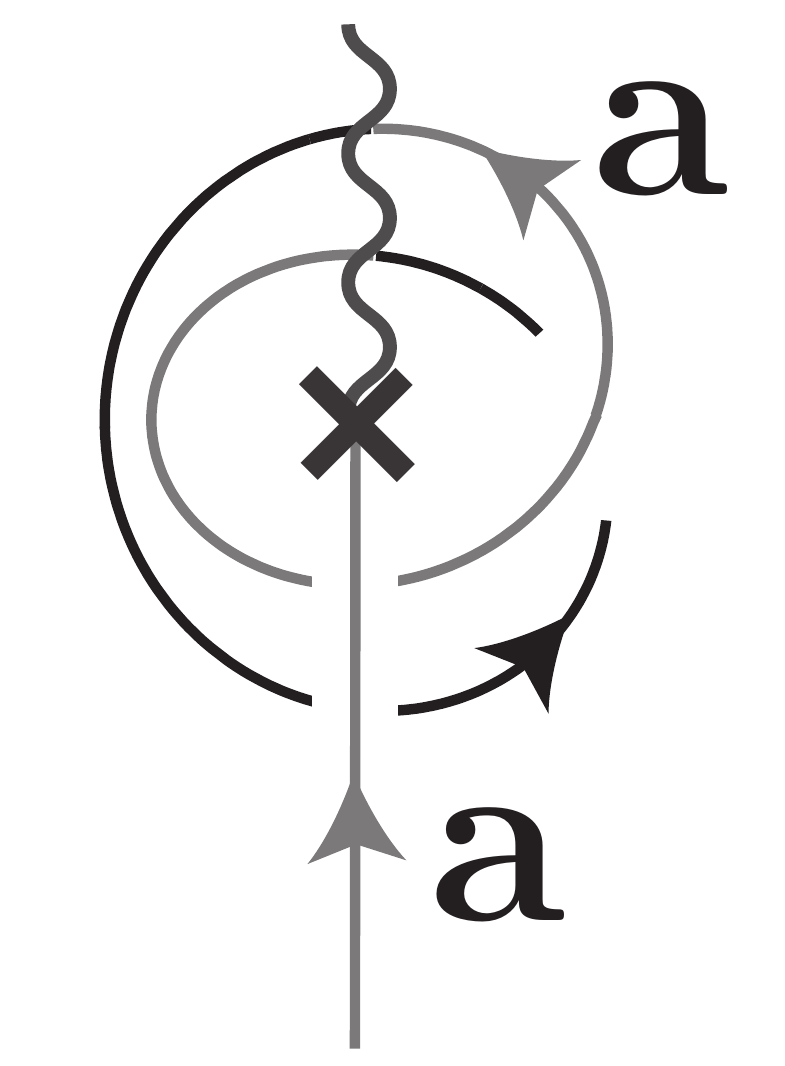}}}=\theta_{\bf a}\Theta^0_{\bf a}\vcenter{\hbox{\includegraphics[width=0.17in]{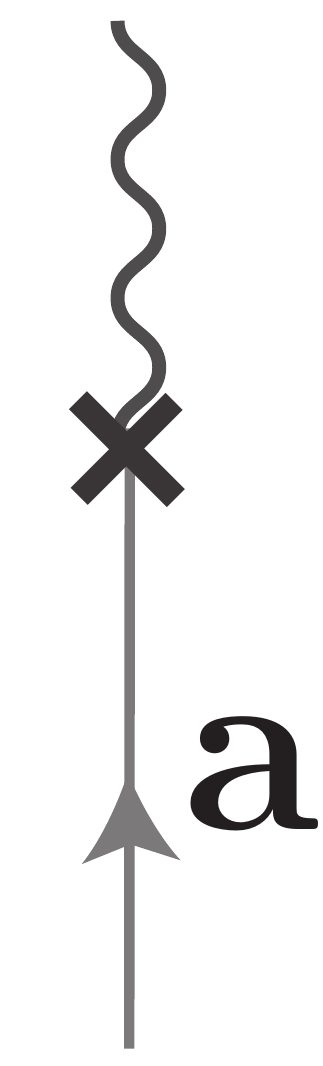}}}=e^{\frac{2\pi i}{m+n}\lambda\left(\frac{\lambda+m+n}{2}\right)}\vcenter{\hbox{\includegraphics[width=0.17in]{Z2spin3.pdf}}}\label{720rotation}\end{align} We notice that the ribbon (or ``pair of pants") identity for braiding a pair of defects is violated $R^{\sigma_{\lambda_1}\sigma_{\lambda_2}}_{\bf a}R^{\sigma_{\lambda_2}\sigma_{\lambda_1}}_{\bf a}=\theta_{\bf a}$, which is different from $\theta_{\bf a}\theta_{\sigma_{\lambda_1}}^{-1}\theta_{\sigma_{\lambda_2}}^{-1}$ when the defects carry fractional charge and have non-trivial spin. It differentiates semiclassical defects from quantum deconfined non-abelian anyons, and forms an obstruction to a fully braided category unless the bilayer structure is melted.

Defect braiding statistics is encoded by the following average over admissible fusion channel within the defect sector that mimics a modular $S$-matrix. \begin{align}S_{\sigma_{\lambda_1}\sigma_{\lambda_2}}=\frac{1}{\mathcal{D}_{\sigma}}\vcenter{\hbox{\includegraphics[width=0.5in]{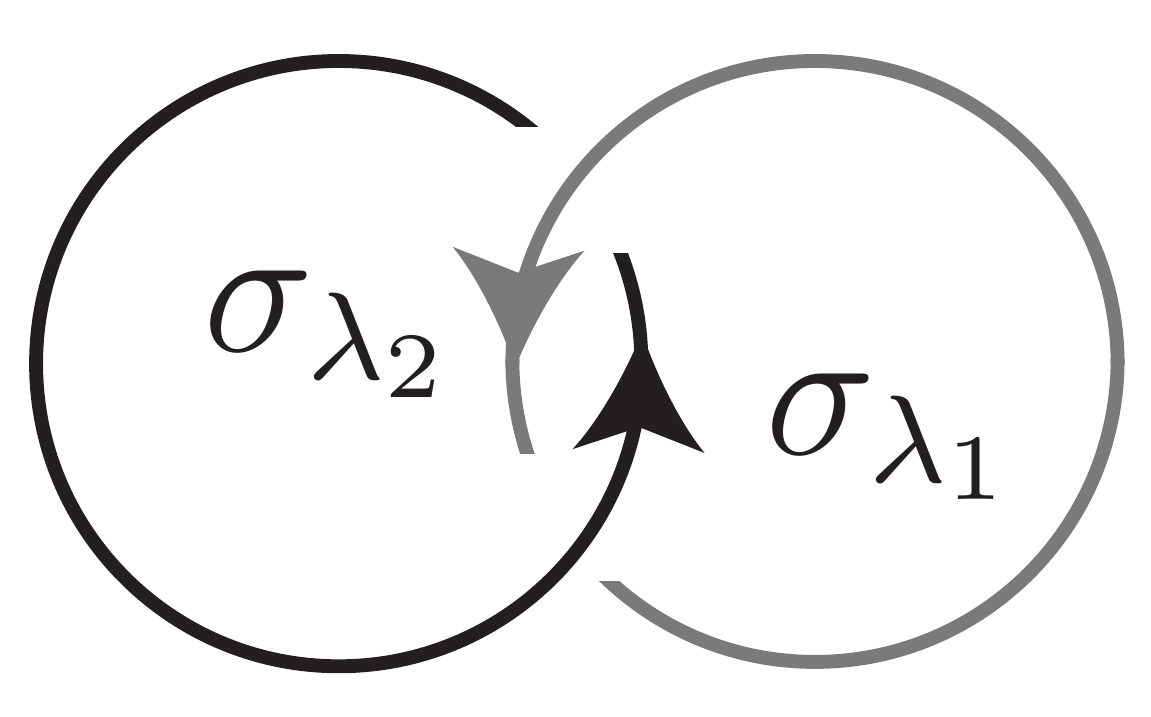}}}=\frac{1}{\mathcal{D}_{\sigma}}\sum_{\bf a}R^{\overline{\sigma_{\lambda_2}}\sigma_{\lambda_1}}_{\bf a}R^{\sigma_{\lambda_1}\overline{\sigma_{\lambda_2}}}_{\bf a}\label{defectSmatrix}\end{align} where $\mathcal{D}_\sigma=\left(\sum_\lambda d_{\sigma_\lambda}^2\right)^{1/2}=\left|m^2-n^2\right|^{1/2}$ is the total quantum dimension of the defect sector, and $\overline{\sigma_\lambda}=\sigma_{-\lambda}$ is the anti-particle of $\sigma_\lambda$. The Gaussian sum has closed form solution \begin{align}&S_{\sigma_{\lambda_1}\sigma_{\lambda_2}}=\frac{z_0}{\sqrt{m+n}}e^{\frac{2\pi i}{m+n}(2^{-1})^2(\lambda_1-\lambda_2)^2}\label{braidingSmatrix}\end{align} where $2^{-1}=(m^2-n^2+1)/2$, and the scalar $z_0$ is 1 if $|m-n|\equiv1$ mod 4, $i(-1)^{sgn(m-n)}$ if $|m-n|\equiv3$ mod 4, or $e^{\frac{i\pi}{4}sgn(m-n)}[1+(-1)^{\lambda_1+\lambda_2+(m-n)/2}]/\sqrt{2}$ if $n$ even.

\begin{figure}[ht]
\includegraphics[width=3in]{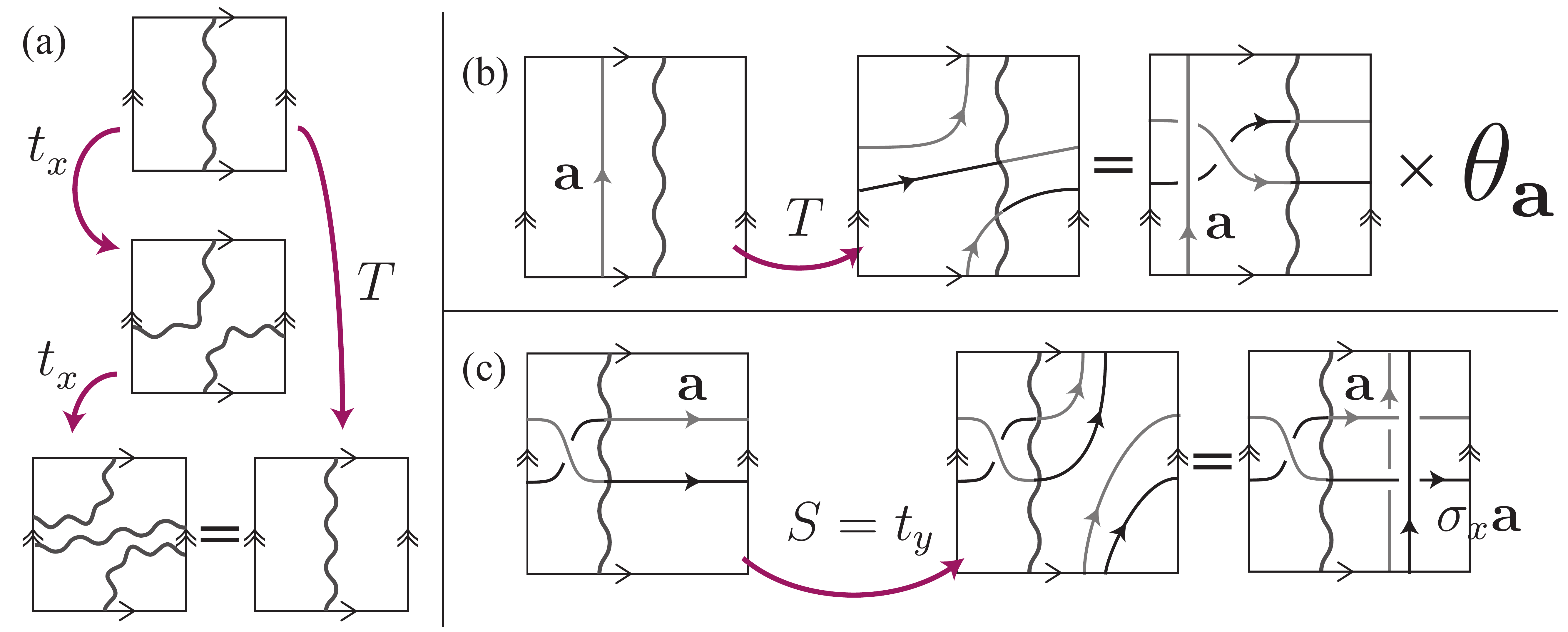}
\caption{Dehn twists. (a) Double Dehn twist $T=t_x^2$ along the horizontal direction that leaves the branch cut (wavy line) invariant. (b) $T$-action on a vertical Wilson loop $\hat{W}_{\bf a}$. (c) Dehn twist $S=t_y$ along the vertical (longitudinal) direction and its action on a double Wilson loop $\hat{\Theta}_{\bf a}$.}\label{fig:dehntwist}
\end{figure}
Defect braiding $S$ and exchange $T$ matrices can alternatively be evaluated as geometric {\em congruent} transformation of ground states on a torus with a twofold branch cut, which occupies a particular cycle, say the $y$-direction, on the torus (see figure~\ref{fig:dehntwist}). The subgroup of modular transformations that leaves the branch cut invariant [i.e. fixes the $y$-cycle with $\mathbb{Z}_2$ coefficient in $H_1(T^2;\mathbb{Z}_2)=\mathbb{Z}_2^2$] is known as a congruent subgroup \begin{align}\Gamma_0(2)=\left\langle\begin{array}{*{20}c}S=t_y\\T=t_x^2\end{array}\left|\begin{array}{*{20}c}(ST^{-1})^2=C,C^2=1\\SCS^{-1}=TCT^{-1}=C\end{array}\right.\right\rangle\end{align} where $t_x,t_y$ are Dehn twists on the torus along the $x,y$ direction. The Wilson operator algebra is generated by quasiparticle strings $\hat{W}_{\bf a}$ of moving ${\bf a}$ along the vertical cycle and $\hat{\Theta}_{\bf a}$ twice along the horizontal cycle (see figure~\ref{fig:dehntwist}). $\hat{W}_{\bf a}$ corresponds the quasiparticle string attaching to a twofold defect and gives rise to its species $\lambda={\bf a}\cdot{\bf t}$, and $\hat{\Theta}_{\bf a}$ is equivalent to the double Wilson operator in figure~\ref{fig:doubleloop} that measures the species label. As these operator can pass across the branch cut and change its label, they obey the twofold symmetry $\hat{W}_{\bf a}=\hat{W}_{\sigma_x{\bf a}}$, $\hat{\Theta}_{\bf a}=\hat{\Theta}_{\sigma_x\bf a}$ and satisfy the fusion relations $\hat{W}_{{\bf a}+{\bf b}}=\hat{W}_{\bf a}\hat{W}_{\bf b}$, $\hat{\Theta}_{{\bf a}+{\bf b}}=e^{2\pi i{\bf a}^TK^{-1}\sigma_x{\bf b}}\hat{\Theta}_{\bf a}\hat{\Theta}_{\bf b}$.
They commute up to the braiding phase of intersection $\hat{\Theta}_{\bf a}\hat{W}_{\bf b}=e^{\frac{2\pi i}{m+n}({\bf a}\cdot{\bf t})({\bf b}\cdot{\bf t})}\hat{W}_{\bf b}\hat{\Theta}_{\bf a}$ and transform under the congruent generators $S$ and $T$ according to $\hat{T}\hat{W}_{\bf a}\hat{T}^\dagger=\theta_{\bf a}\hat{W}_{\bf a}\hat{\Theta}_{\bf a}$ and $\hat{S}\hat{\Theta}_{\bf a}\hat{S}^\dagger=\hat{W}_{\bf a}\hat{\Theta}_{\bf a}\hat{W}_{\sigma_x\bf a}$ (see figure~\ref{fig:dehntwist}) while keeping $\hat{T}\hat{\Theta}_{\bf a}\hat{T}^\dagger=\hat{\Theta}_{\bf a}$ and $\hat{S}\hat{W}_{\bf a}\hat{S}^\dagger=\hat{W}_{\bf a}$.

The $|m+n|$-fold degenerate ground states are simultaneous eigenstates of $\hat{\Theta}_{\bf a}$ and $\hat{T}$ that corresponds defect species, $\hat{\Theta}_{\bf a}|\lambda\rangle=\Theta^\lambda_{\bf a}|\lambda\rangle$, $\hat{W}_{\bf a}|\lambda\rangle=|\lambda+{\bf a}\cdot{\bf t}\rangle$, where the eigenvalue $\Theta^\lambda_{\bf a}$ is given by \eqref{doubleloopphase}. \begin{align}\hat{T}|\lambda\rangle&=\hat{T}\hat{W}_{\bf a}|0\rangle=\theta_{\bf a}\hat{W}_{\bf a}\hat{\Theta}_{\bf a}\hat{T}|0\rangle=T_0\theta_{\bf a}\Theta^0_{\bf a}|\lambda\rangle\end{align} where ${\bf a}$ is any quasiparticle with charge $\lambda={\bf a}\cdot{\bf t}$, and $T_0$ is the eigenvalue $\langle0|\hat{T}|0\rangle$. And therefore up to a constant phase, the congruent $T$-matrix that represents double Dehn twist $t_x^2$ is diagonal with spin entries \eqref{exchangespin}, $T_{\lambda_1\lambda_2}\propto\delta_{\lambda_1\lambda_2}\theta_{\lambda_1}$. The commuting $\hat{W}_{\bf a}$ and $\hat{S}$ share the simultaneous eigenstate $|\emptyset\rangle=\frac{1}{\sqrt{|m+n|}}\sum_\lambda|\lambda\rangle$, which generates all ground states by $|\lambda\rangle=\frac{1}{\sqrt{|m+n|}}\hat{W}_{\bf a}{\sum_{\bf b}}'(\Theta^0_{\bf b})^{-1}\hat{\Theta}_{\bf b}|\emptyset\rangle$, where $\lambda={\bf a}\cdot{\bf t}$ and the sum is restricted to any collection of $|m+n|$ quasiparticles ${\bf b}$ with distinct fractional charges. The congruent $S$-matrix that represents Dehn twist $t_y$ has entries \begin{align}\langle\lambda_1|\hat{S}|\lambda_2\rangle&=\frac{S_\emptyset}{|m+n|}{\sum_{\bf b}}'\langle\emptyset|\hat{\Theta}_{\bf b}^\dagger\hat{W}_{{\bf a}_2-{\bf a}_1+{\bf b}}\hat{\Theta}_{\bf b}|\emptyset\rangle\nonumber\\&=\frac{S_\emptyset}{|m+n|}\sum_{l=0}^{|m+n|-1}e^{\frac{2\pi i}{m+n}l(\lambda_2-\lambda_1-l)}\label{Sgaussiansum}\end{align} where $S_\emptyset$ is the eigenvalue $\langle\emptyset|\hat{S}|\emptyset\rangle$. After evaluating the Gaussian sum, \eqref{Sgaussiansum} equals the braiding $S$-matrix \eqref{braidingSmatrix} up to a constant phase. The braiding $S$ and exchange spin $T$ matrices in \eqref{braidingSmatrix} and \eqref{exchangespin} generate a unitary representation for the congruent subgroup $\Gamma_0(2)$ where \begin{align}C_{\lambda_1\lambda_2}\equiv\left[(ST^\dagger)^2\right]_{\lambda_1\lambda_2}=sgn(m+n)\delta_{\lambda_1,-\lambda_2}\end{align} serves as a charge conjugation operator to defect species.

{\em Acknowledgements} This work was supported by the Simons Foundation (JT) and National Science Foundation through grant DMR 09-03291 (AR) and DMR-1064319 (XC).


\newpage
\appendix
\onecolumngrid
\section{Supplementary Materials on Evaluating the defect \texorpdfstring{$F$}{F} and \texorpdfstring{$R$}{R}-symbols}
Quasiparticles of the $(mmn)$-FQH state live on the quotient lattice $\mathcal{A}=\mathbb{Z}^2/K$, where the $K$-matrix is given in \eqref{Kmatrix}. When $m=0$, the theory has the same anyon content as the $\mathbb{Z}_n$ Kitaev toric code and the quotient lattice decomposes into $\mathcal{A}=\mathbb{Z}_n\oplus\mathbb{Z}_n$ as all quasiparticles can be uniquely be written as ${\bf a}=a_1{\bf e}+a_2{\bf m}$ where ${\bf e}=(1,0)$ and ${\bf m}=(0,1)$. When $n$ is odd, 2 is invertible in $\mathbb{Z}_{m^2-n^2}$ by $2^{-1}=(m^2-n^2+1)/2$ and the transformation $W=\sigma_z+\sigma_x\in SL(2;\mathbb{Z}_{m^2-n^2})$ gives an isomorphism of the quotient lattice $\mathcal{A}\to\mathcal{A}$ with inverse $W^{-1}=2^{-1}W$ as $WW^{-1}=W^{-1}W=(m^2-n^2+1)\openone\equiv\openone$ mod $K$. Thus $\mathcal{A}=\mathbb{Z}_{m+n}\oplus\mathbb{Z}_{m-n}$ where each quasiparticle has a unique decomposition ${\bf a}=x{\bf t}+\alpha{\bf s}$ where ${\bf t}=(1,1)$ and ${\bf s}=(1,-1)$ are the charge and bilayer/spin vectors. The FQH quasiparticle $R$-symbols are chosen to be multilinear homomorphisms $\mathcal{A}\times\mathcal{A}\to\mathbb{Z}_{m^2-n^2}$. (i) $R^{{\bf a}{\bf b}}=e^{\frac{2\pi i}{n}a_2b_1}$ for $\mathbb{Z}_n$ toric code when $m=0$. (ii) $R^{{\bf a}{\bf b}}=e^{2\pi i2^{-1}{\bf a}^TK^{-1}{\bf b}}$ for $(mmn)$-FQH states with $n$ odd. Under this convention, $F$-symbols can be chosen to be trivial one $F^{{\bf a}{\bf b}{\bf c}}_{{\bf a}+{\bf b}+{\bf c}}=1$ and the quasiparticle strings satisfy the crossing, bootstrap and bending rules. \begin{align}\vcenter{\hbox{\includegraphics[width=0.35in]{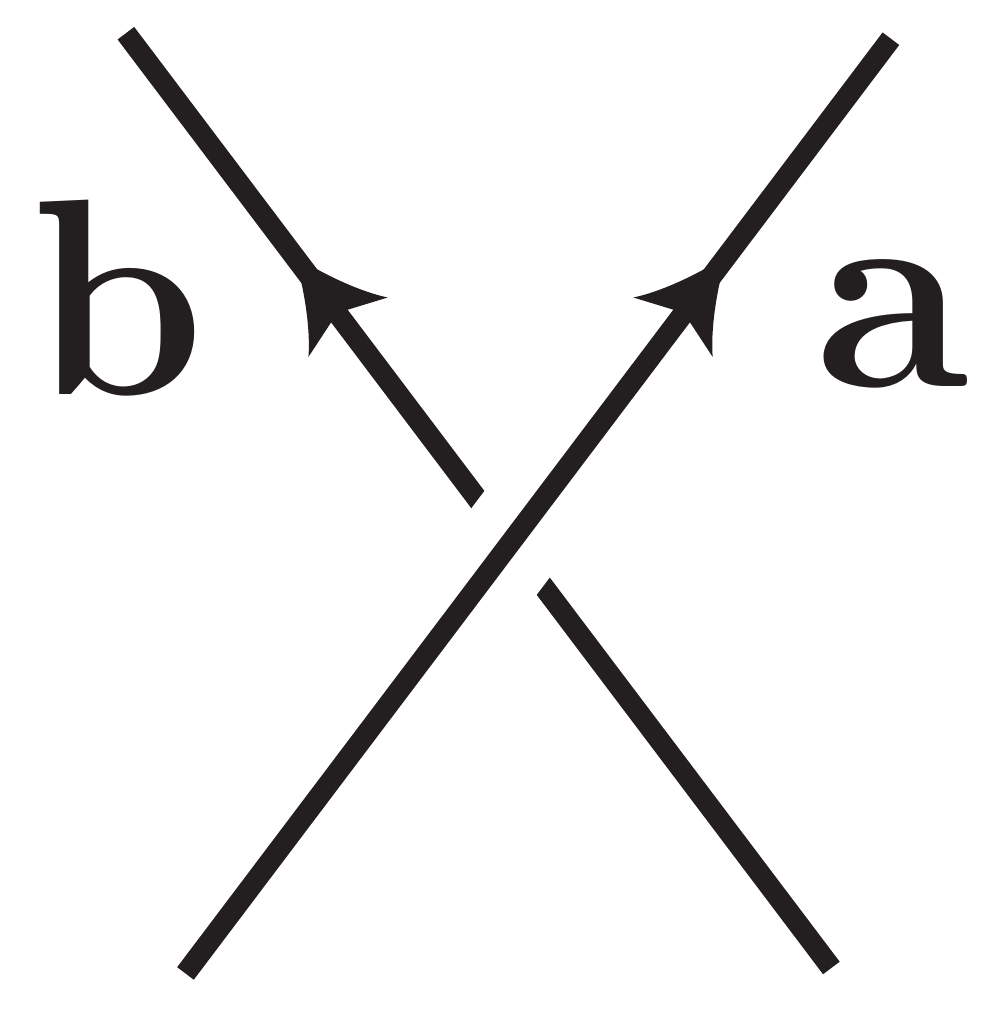}}}=\vcenter{\hbox{\includegraphics[width=0.35in]{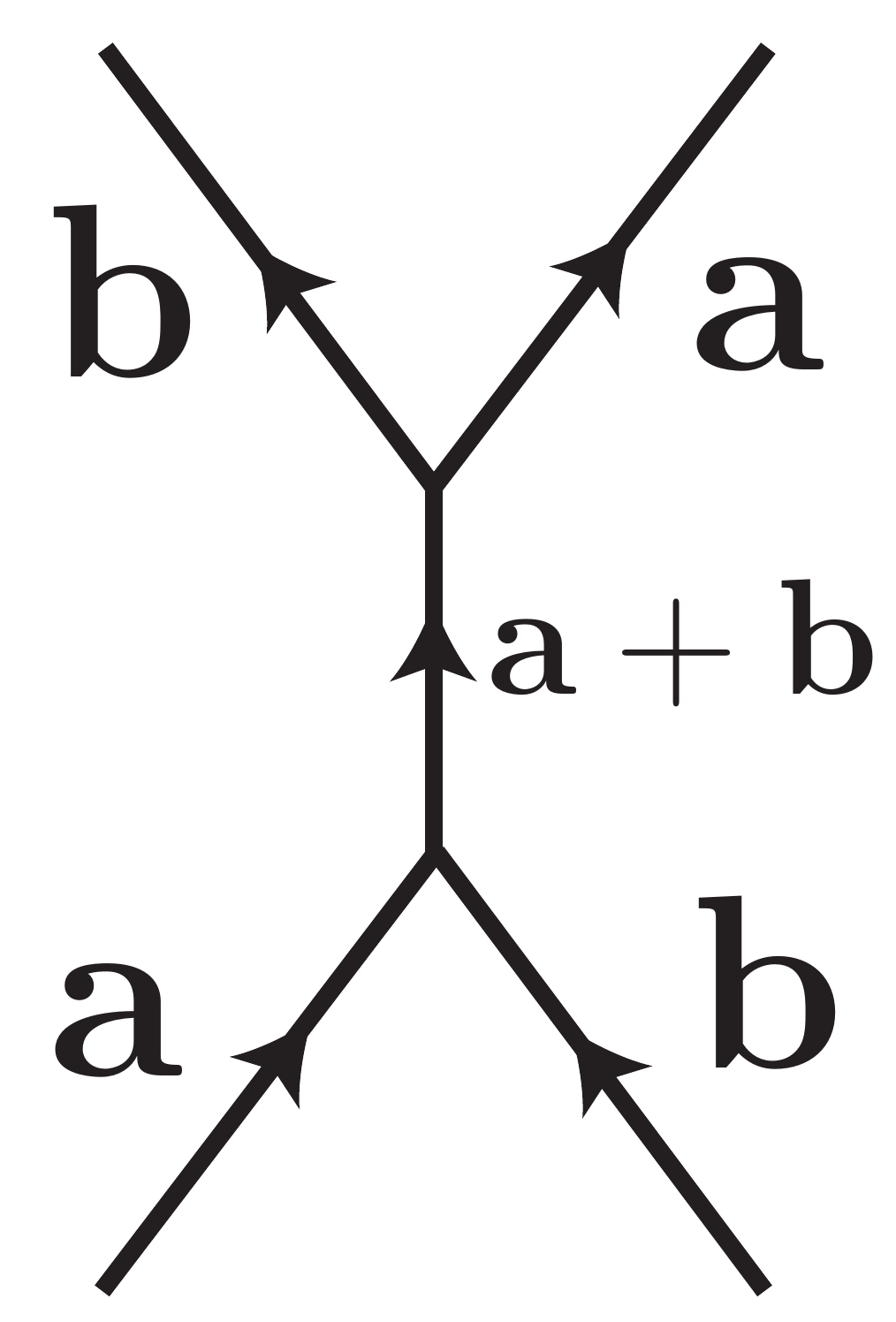}}}R^{{\bf a}{\bf b}}=\vcenter{\hbox{\includegraphics[width=0.35in]{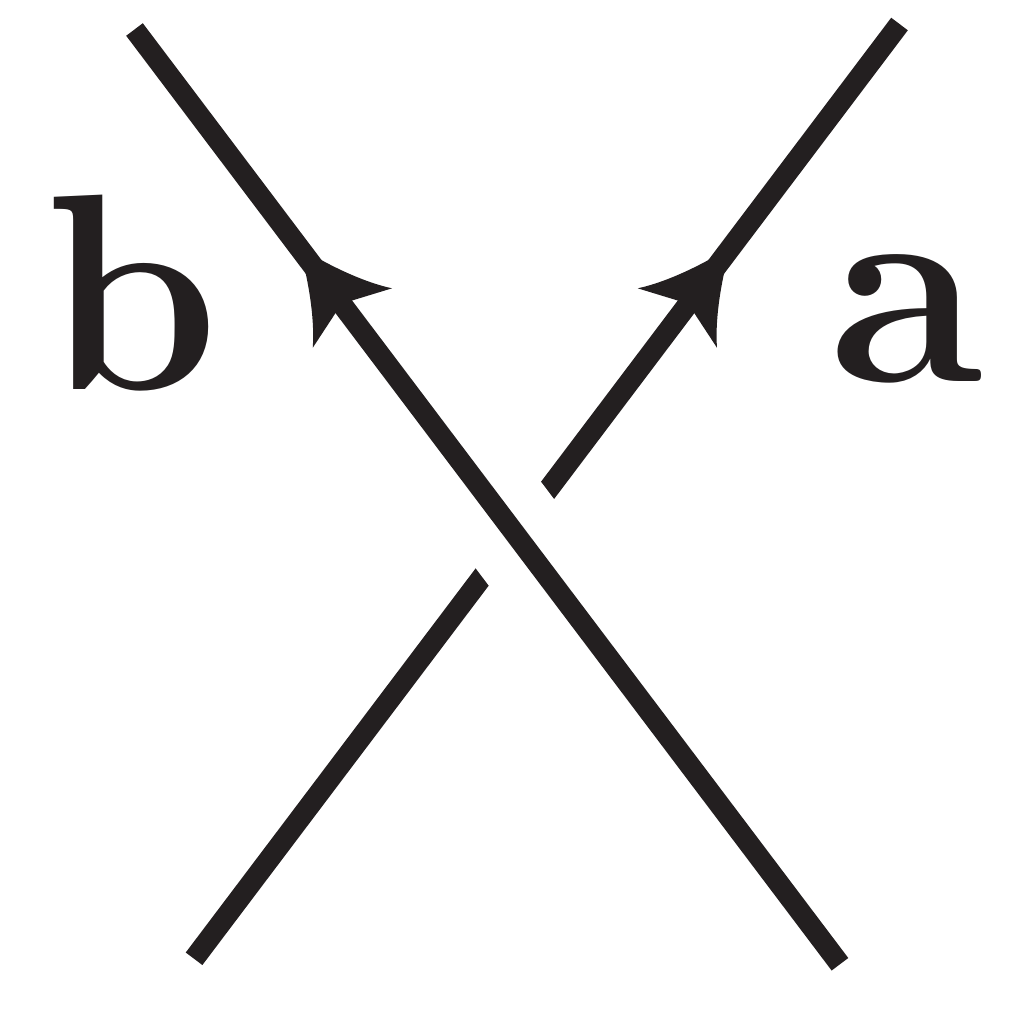}}}R^{{\bf a}{\bf b}}R^{{\bf b}{\bf a}},\quad\vcenter{\hbox{\includegraphics[width=0.45in]{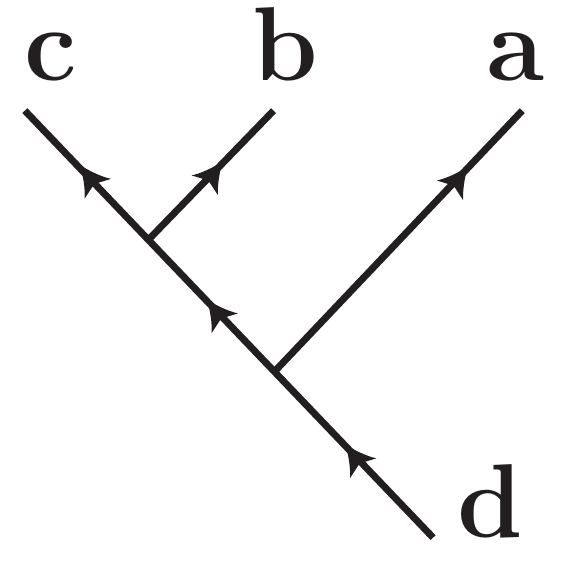}}}=\vcenter{\hbox{\includegraphics[width=0.45in]{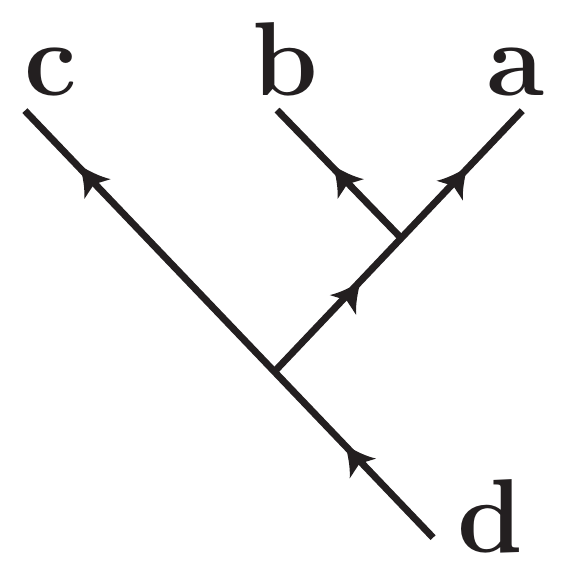}}},\quad\vcenter{\hbox{\includegraphics[width=0.35in]{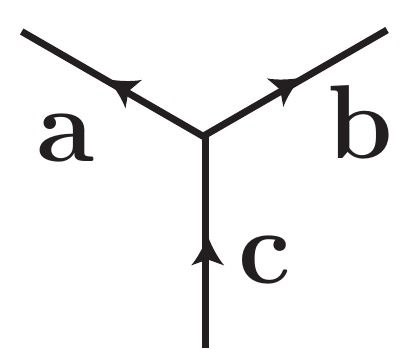}}}=\vcenter{\hbox{\includegraphics[width=0.35in]{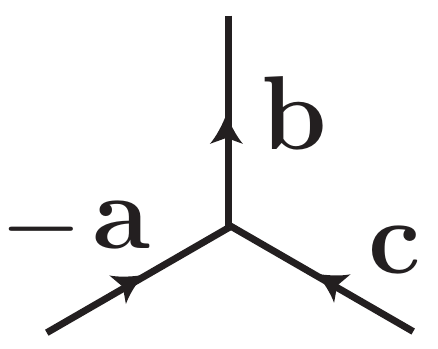}}}=\vcenter{\hbox{\includegraphics[width=0.35in]{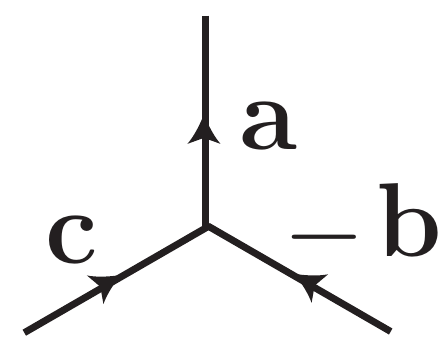}}}\end{align} 

The double loop operator $\hat{\Theta}_{\mathbf{a}}$ carries an anyon $\mathbf{a}$ twice around a twist defect as shown in fig. We have the relations (i) $\hat{\Theta}_{\mathbf{a}} = \hat{\Theta}_{\sigma_x\mathbf{a}}$, (ii) $\hat{\Theta}_{\mathbf{a+b}} = \hat{\Theta}_{\mathbf{a}}\hat{\Theta}_{\mathbf{b}}e^{2\pi
i \mathbf{a}^T K^{-1}\sigma_x\mathbf{b}}$ and (iii) $(\Theta^0_{\mathbf a})^2 = e^{2\pi
i \mathbf{a}^T K^{-1}\sigma_x\mathbf{b}}$, where $\Theta^0_{\mathbf a}$ is the eigenvalue for the bare defect ($\lambda=0$). $\Theta^\lambda_{\mathbf{a}}=c\exp\left[2\pi i {\bf a}^T K^{-1}\sigma\left(\frac{1}{2}{\bf a}+\lambda{\bf t}\right)\right]$ satisfies (i--iii), where the constant $c=\pm 1$ is fixed by requiring that $\hat{\Theta}_{\mathbf{a}+K\mathbf{v}}=\hat{\Theta}_{\mathbf{a}}$, so that $\hat{\Theta}$ is well-defined on the anyon lattice. The result is eq.\eqref{doubleloopphase}.

Basis transformation between different fusion trees with defect channels can be generated by a consistent set of fundamental $F$-moves. In table~\ref{tab:Fsymbols}, we list all the $F$-symbols for $\mathbb{Z}_n$ toric code and $(mmn)$-FQH states. These $F$-symbols are calculated by gluing the splitting states defined in figure~\ref{fig:splittingstates} by joining quasiparticle strings. Here we illustrate the calculation of the $F$ symbol that relates $(\sigma_{\lambda_3}\times\sigma_{\lambda_2})\times\sigma_{\lambda_1}\to\sigma_{\lambda_3}\times(\sigma_{\lambda_2}\times\sigma_{\lambda_1})$ in $(mmn)$-FQH states for $n$ odd. 
\begin{figure}[ht]
\includegraphics[width=6.5in]{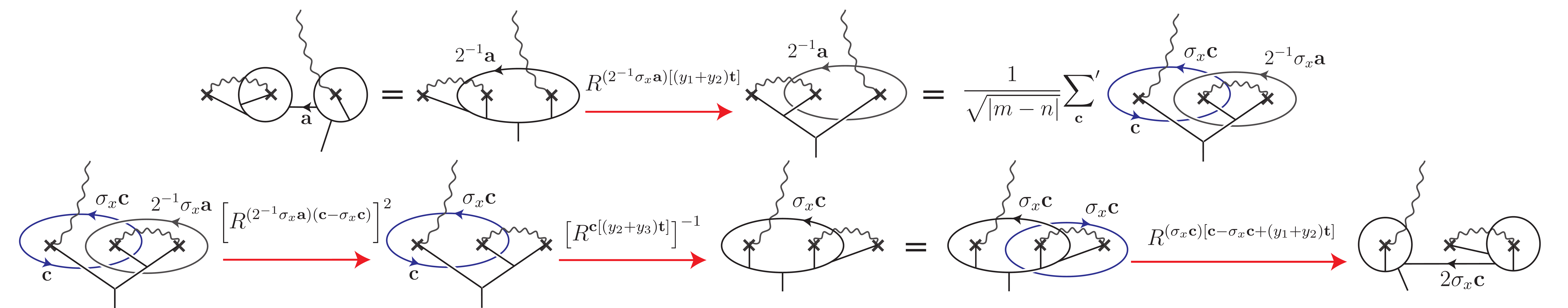}
\caption{Derivation of the $F$-matrix $\left[F^{\sigma_{\lambda_3}\sigma_{\lambda_2}\sigma_{\lambda_1}}_{\sigma_{\lambda_1+\lambda_2+\lambda_3}}\right]_{\bf a}^{\bf b}$. The sum over Wilson loops $\hat{W}_{\bf c}$ (blue) is restricted to an arbitrary set of $|m-n|$ quasiparticles ${\bf c}=z{\bf t}+\gamma\bf s{}$ with distinct spin components $\gamma$ as $\hat{W}_{\bf c}|GS\rangle=\hat{W}_{\sigma_x{\bf c}}^\dagger|GS\rangle$ (see figure~\ref{fig:doubleloop}). The operator $\frac{1}{\sqrt{|m-n|}}{\sum_{\bf c}}'\hat{W}_{\bf c}$ flips the branch cut configuration from joining $\sigma_{\lambda_3}\times\sigma_{\lambda_2}$ to connecting $\sigma_{\lambda_2}\times\sigma_{\lambda_1}$.}\label{fig:Fsss}
\end{figure}
The splitting state for $|V^{\sigma_{\lambda_3}\sigma_{\lambda_2}}_{\bf a}\rangle\otimes|V^{{\bf a}\sigma_{\lambda_1}}_\sigma\rangle$ is represented by the first diagram and that for $|V^{\sigma_{\lambda_3}\bf b}_{\sigma}\rangle\otimes|V^{\sigma_{\lambda_2}\sigma_{\lambda_1}}_{\bf b}\rangle$ is at the last by setting ${\bf b}=2\sigma_x{\bf c}$. The $F$-symbol is given by the product of all the crossing phases and the normalization factor $1/\sqrt{|m-n|}$. The basis transformation between any two fusion trees can be achieved by a series of $F$-symbols, and the result is independent from the choice of moves. This is ensured by the pentagon identity $FF=FFF$ and the MacLane's coherence theorem~\cite{Kitaev06}. $F$-symbols listed in table~\ref{tab:Fsymbols} can be shown to satisfy the $2^4=16$ identities.

The $R$-symbol for exchanging a pair of defects is evaluated by rotating the splitting space for twist defect in figure~\ref{fig:splittingstates}(c). We illustrate this for the case when $n$ is odd. Eq.\eqref{Rsymbol} is a product of crossing phases and a double Wilson operator.
\begin{figure}[ht]
\includegraphics[width=6.5in]{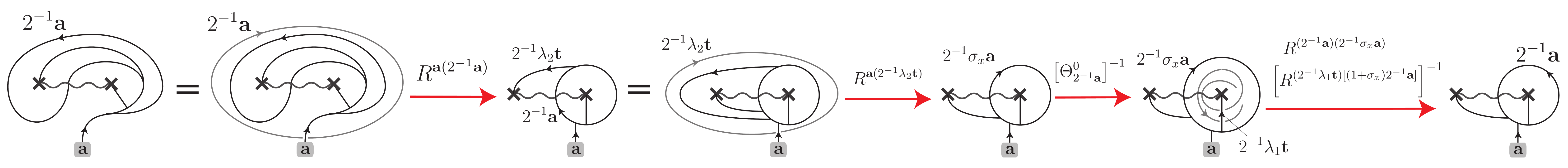}
\end{figure}


\end{document}